\documentclass[usenatbib]{mn2e} 
\usepackage{amsmath} 
\usepackage{amssymb} 
\usepackage{graphics}
\usepackage{graphicx}
\usepackage{epsfig}  
\def\be{\begin{equation}}
\def\ee{\end{equation}}
\def\ba{\begin{eqnarray}}
\def\ea{\end{eqnarray}}

\usepackage{color}
\definecolor{red}{rgb}{1,0.0,0.0}

\definecolor{blue}{rgb}{0.1,0.3,0.9}

\usepackage[normalem]{ulem}
\definecolor{darkgreen}{rgb}{0.0,0.5,0.0}

\newcommand{\beq}{\begin{eqnarray}}  
\newcommand{\eeq}{\end{eqnarray}}  
 
\newcommand{\apj}{ApJ}  
\newcommand{\apjs}{ApJS}  
\newcommand{\apjl}{ApJL}  
  
\newcommand{\mnras}{MNRAS}  
  
\newcommand{\aap}{A\&A}  
\newcommand{\aaps}{A\&AS}  
\newcommand{\araa}{ARA\&A}  
\newcommand{\nat}{Nature}  
\newcommand{\physrep}{PhR}

\newcommand{\ly}{{\ifmmode{{\rm Ly}\alpha}\else{Ly$\alpha$}\fi}}
\newcommand{\hMpc}{{\ifmmode{h^{-1}{\rm Mpc}}\else{$h^{-1}$Mpc }\fi}}  
\newcommand{\hGpc}{{\ifmmode{h^{-1}{\rm Gpc}}\else{$h^{-1}$Gpc }\fi}}  
\newcommand{\hmpc}{{\ifmmode{h^{-1}{\rm Mpc}}\else{$h^{-1}$Mpc }\fi}}  
\newcommand{\hkpc}{{\ifmmode{h^{-1}{\rm kpc}}\else{$h^{-1}$kpc }\fi}}  
\newcommand{\hMsun}{{\ifmmode{h^{-1}{\rm {M_{\odot}}}}\else{$h^{-1}{\rm{M_{\odot}}}$}\fi}}  
\newcommand{\hmsun}{{\ifmmode{h^{-1}{\rm {M_{\odot}}}}\else{$h^{-1}{\rm{M_{\odot}}}$}\fi}}  
\newcommand{\Msun}{{\ifmmode{{\rm {M_{\odot}}}}\else{${\rm{M_{\odot}}}$}\fi}}  
\newcommand{\msun}{{\ifmmode{{\rm {M_{\odot}}}}\else{${\rm{M_{\odot}}}$}\fi}}

\newcommand{\rand}{{\ifmmode{{\mathcal{R}}}\else{${\mathcal{R}}$ }\fi}}  
\newcommand{\hs}{{\hspace{1mm}}}  

\def\lsim{~\rlap{$<$}{\lower 1.0ex\hbox{$\sim$}}}

\def\gsim{~\rlap{$>$}{\lower 1.0ex\hbox{$\sim$}}}

\begin{document}

\title[Stochasticity in Dwarf Galaxies during Reionization]{Effects of Star Formation Stochasticity
  on the Ly$\alpha$ \& Lyman Continuum Emission from Dwarf Galaxies during Reionization}
\author[J.E. Forero-Romero and M. Dijkstra]{
\parbox[t]{\textwidth}{\raggedright 
  Jaime E. Forero-Romero$^{1}$ \thanks{IAU Gruber Fellow, Email: forero@berkeley.edu} and Mark Dijkstra$^{2}$
}
\vspace*{6pt}\\
$^{1}$Department of Astronomy, University of California, Berkeley, CA 94720-3411, USA\\
$^{2}$Max-Planck Institut fuer Astrophysik, Karl-Schwarzschild-Str. 1, 85741 Garching, Germany
}
\maketitle

\begin{abstract}
Observations of distant galaxies play a key role in improving our understanding of the Epoch of Reionization (EoR). The observed Ly$\alpha$ emission line strength -- quantified by its restframe equivalent width (EW) -- provides a valuable diagnostic of stellar populations and dust in galaxies during and after the EoR. In this paper we quantify the effects of star formation stochasticity on the predicted Ly$\alpha$ EW in dwarf galaxies, using the publicly available code SLUG (used to `Stochastically Light Up Galaxies'). We compute the number of hydrogen ionizing photons, as well as flux in the Far UV for a set of models with star formation rates (SFR) in the range $10^{-3}$-$1$ \Msun\hs yr$^{-1}$. From these fluxes we compute the luminosity, $L_{\alpha}$, and the EW of the Ly$\alpha$ line. We find that stochasticity alone induces a broad distribution in $L_{\alpha}$ and EW at a fixed SFR, and that the widths of these distributions decrease with increasing SFR. We parameterize the EW probability density function (PDF) as an SFR--dependent double power law. We find that it is possible to have EW as low as $\sim$EW$_{0}/4$ and as high as $\sim 3\times$EW$_{0}$, where EW$_{0}$ denotes the expected EW in the absence of stochasticity. We argue that stochasticity may therefore be important when linking drop-out and narrow-band selected galaxies, when identifying population III galaxies, and that it may help to explain the large EW (EW$\gsim 100-200$ \AA) observed for a fraction of Ly$\alpha$ selected galaxies. Finally, we show that stochasticity can also affect the inferred escape fraction of ionizing photons from galaxies. In particular, we argue that stochasticity may simultaneously explain the observed anomalous ratios of the Lyman continuum flux density to the (non-ionizing) UV continuum density in so-called Lyman-Bump galaxies at $z=3.1$, as well as the absence of such objects among a sample of $z=1.3$ drop-out galaxies.
\end{abstract}
\begin{keywords}
galaxies: high-redshift - galaxies: star formation - line: formation
\end{keywords}

\section{Introduction}
\label{sec:introduction}
The Epoch of Reionization (EoR) represents a milestone in the evolution of our Universe, during which hydrogen gas in most of its volume changed from fully neutral to fully ionized. Observations of the cosmic microwave background support that reionization started at redshift
$z\sim 20$ while quasars observations demonstrate that the process was completed
at $z\sim 5-6$ \citep[e.g.][]{Mesinger10,Pritchard10}.  One key to new insights in this area are observations of high redshift galaxies, as they will allow for the construction of a complete physical picture of the reionization process.

Observations of distant Lyman Break Galaxies (LBGs) and Lyman $\alpha$
Emitters (LAEs) in the redshift range $4<z<10$ have provided the most prominent windows on the star formation history at these early epochs \citep[see e.g.][and references therein]{Robertson10}. LBGs are galaxies that have been selected using broad-band surveys, which are sensitive to the `Lyman break' in the spectra of star forming galaxies. This break is a consequence of typical effective temperatures of O \& B stars, and of interstellar and intergalactic absorption of flux blueward of the Ly$\alpha$ wavelength \citep{Steidel96}. Surveys for LBGs (or `drop-out' galaxies) are most sensitive to UV rest-frame continuum emission of galaxies. LAEs are galaxies that have been selected from narrow-band surveys which are sensitive to high-redshift galaxies that have strong Ly$\alpha$ line emission \citep[e.g.][]{Rhoads00}. The Ly$\alpha$ line originates in HII regions that surrounded O \& B stars. As a result, the Ly$\alpha$ line flux (and strength) in LAEs are more directly connected to the ionizing photon budget \citep[e.g.][]{2003A&A...397..527S}.

The `strength' of the  Lyman $\alpha$ emission line is quantified by its rest-frame equivalent width (EW), which corresponds to the
ratio between the line intensity and the continuum. The Ly$\alpha$ EW is an important physical quantity associated with galaxies during the EoR for (at least) three reasons : ({\it i}) narrowband surveys for LAEs are typically sensitive to galaxies for which the Ly$\alpha$ EW exceeds some threshold value \citep[which can range between EW$\sim 20-65$ \AA, see e.g.][]{Ouchi08}. The Ly$\alpha$ EW is clearly an important physical quantity that links the LBG and LAE populations; ({\it ii}) the `first' galaxies that formed from pristine gas are expected to be characterised by unusually large EW (well in excess of EW$\sim 200$ \AA), which may be one of their most prominent observational signatures \citep{2003A&A...397..527S,2009MNRAS.399...37J,Raiter10}; ({\it iii}) recent observations of LAEs \& LBGs in the redshift range $z=3-7$ have revealed an apparent non-monotonic evolution in the EW distribution, with significantly less Ly$\alpha$ being observed from galaxies beyond redshift $z>6$ \citep{Stark11a,Stark11b,Pentericci11,Ono12,Schenker12}.

The redshift evolution in the EW distribution at $z>6$ is likely to be connected with the reionization process rather than with secular evolution process in the galaxies \citep[][also see \cite{Shimizu11}]{Forero12}. 
Although current observations do not allow for robust conclusions, analyzing EW distributions provides a powerful technique to probe the population of star forming galaxies -- as well as the ionization state of the IGM -- during and after reionization has been completed \citep[e.g.][also see \cite{Dayal2012}]{Dijkstra11}.


Recently, \citet{2011ApJ...741L..26F} showed that at low levels of star formation (SFR $< 1 M_{\odot}$ yr$^{-1}$), the effects of the statistical description of the stellar Initial
Mass Function (IMF) and the Cluster Mass Function (CMF) emerge. Among the expected effects -- gathered under the name of `stochasticity' -- the most relevant for our discussion is that the flux ratio between ionizing and Far UV photons fluctuates around the mean expected value. \citet{2011ApJ...741L..26F} show how the clustering of star formation together with the sampling of the Initial Mass Function (IMF) cause the ratio of the H$\alpha$ flux and FUV flux density to fluctuate around the mean significantly for local dwarf galaxies. This effect should be also noticeable in the ratio of Ly$\alpha$ flux to FUV flux density i.e. the EW. In other words, stochasticity may affect the {\it intrinsic} Ly$\alpha$ EW-PDFs of faint LBGs ($-18 \lsim M_{\rm UV} \lsim -13$). Because Ly$\alpha$ EW plays an important role in understanding high--redshift galaxy populations and constraining the reionization process, it is important to understand quantitatively the impact of stochasticity.

Furthermore, observational constraints on the `escape fraction' of ionizing photons from star forming galaxies are derived from the observed ratio of LyC flux density to FUV flux density, which is subject to the clustering of star formation and stochastic sampling of the IMF, especially at SFR$\lsim 1\hs M_{\odot}$ yr$^{-1}$. It has long been realized that such galaxies can provide the bulk of the ionizing photons needed to reionize the IGM and to maintain it ionized \citep[at a fixed escape fraction, see e.g. Fig~31 of][]{Barkana01}. A more recent analysis by \cite{2012arXiv1201.0757K} echoes these findings. \citet{2012arXiv1201.0757K} show that in plausible scenarios for reionization, star forming galaxies with very faint magnitudes $M_{\lim}\sim -13$ (corresponding to SFR $\sim 10^{-2}$\Msun yr$^{-1}$) contribute significantly to the overall budget of ionizing photons. While high--redshift galaxies with such low star formation rates remain too faint to be detected even by future facilities, the observation of galaxies with SFRs$=10^{-2}-1\hs M_{\odot}$ yr$^{-1}$ at lower redshifts $z<4$ can help us to understand the physics of such systems. For instance, the escape fraction of ionizing photons from galaxies with SFR within the upper end of that range have already been constrained by observations at $z\sim 3.0$ \citep[][who constrained the escape fraction down to SFR$\sim 1 M_{\odot}$ yr$^{-1}$]{2009ApJ...692.1287I}. The possible dependence of these estimates on stochasticity provides an additional motivation to study the consequences of the stochastic sampling of the IMF.

In this paper we set out to quantify the effects of stochasticity in
the EW distribution of dwarf Ly$\alpha$ emitting galaxies. We
implement highly simplified models for dwarf galaxies at high redshift
to isolate the stochastic effect. The SFH is constant, single
metallicity, without any extinction effects.  We stress that our objective is not to explain or interpret current observations of LAEs and LBGs, but instead simply to isolate and highlight the impact of stochasticity on predicted values for the Ly$\alpha$ EW.  

This paper is structured as follows. In \S \ref{sec:ew} we summarize the expected effects on the equivalent width stochasticity and present the numerical experiments we perform to quantify this effect. In \S \ref{sec:results} we present the raw results that help us in \S \ref{sec:discussion} to asses the impact of stochasticity in different aspects of the interpretation of observations of LAEs and LBGs at high redshift. Finally, we summarize our conclusions in \S \ref{sec:conclusions}

Throughout this paper we will study the `Lyman continuum' ($\lambda \sim 700$\AA), ionizing continuum ($\lambda<912$\AA), and Far UV (FUV) continuum ($\sim 1500$\AA), where all wavelengths are rest frame. Whenever we refer to UV we imply FUV. Also, the Ly$\alpha$ equivalent widths are measured in the galaxies' restframe.

\section{Stochastic effects in the Equivalent Width}
\label{sec:ew}

To be consistent with observations, we define the
Ly$\alpha$ rest-frame equivalent EW as the ratio between the
intensity of the Ly$\alpha$ line to the flux density redwards of
the line\footnote{Formally, the equivalent width is measured relative to the flux density in the continuum just redward of the Ly$\alpha$ resonance, i.e. at $\lambda \sim 1216 \hs$\AA$+\epsilon$.  The flux density at this wavelength is $\sim (\lambda_{\alpha}/\lambda_{\rm UV})^{\beta}$ times larger than at $\lambda_{\rm UV}=1540$ \AA, where $\beta \sim -2$  \citep{Dunlop12,Fink} denotes the observed slope of the UV continuum at the redshift and UV-magnitudes of interest. Therefore, if we had corrected for the slope in the spectrum of the UV continuum, then we would have found EW$_0\sim 70$ \AA\hs for the \cite{Kennicutt98} star formation calibrators \citep[e.g.][]{westra}.}  

\begin{equation}
{\rm EW} \equiv \frac{L_{Ly\alpha}}{L_{\lambda,UV}} = \frac{\lambda_{\rm UV}}{\nu_{\rm UV}} \frac{L_{Ly\alpha}}{L_{\nu,UV}},
\end{equation}
where subscript $\nu$ or $\lambda$ indicates the flux density in
frequency or wavelength, respectively. The wavelengths and frequencies associated with the Ly$\alpha$ transition and the UV continuum (from the central wavelength for the filter used in SLUG) are $\lambda_{\alpha}=1216$\AA, $\lambda_{\rm UV}=1540$\AA, $\nu_{Ly\alpha}=2.47\times 10^{15}$Hz and $\nu_{\rm UV}=1.97\times 10^{15}$Hz. 

We calculate the intrinsic luminosity of Ly$\alpha$ assuming
that a fraction $0.67$ of the ionizing photons are converted into Ly$\alpha$.
We obtain L$_{{\rm Ly\alpha}}$=$c_{0}\times Q_{H}$ where
$c_{0}=1.04\times 10^{-11}$ erg photon$^{-1}$
\citep{2003A&A...397..527S}. We then have

\begin{equation}
{\rm EW} = \frac{\lambda_{\rm UV}}{\nu_{\rm UV}}\frac{c_{0}Q_{H}}{F_{\nu,{\rm UV}}}.
\label{eq:def}
\end{equation}

In order to compare results on the same grounds results from different
metallicities or star formation rates we introduce a dimensionless
constant 

\begin{equation}
{\mathcal M} \equiv \frac{{\rm EW}}{{\rm EW}_{0}},
\label{eq:M}
\end{equation}
where ${\rm EW}_{0}$ is the equivalent width expected value for
high values of the star formation rate when the stochasticity effects
are negligible. 

We point out that under this convention EW$_{0}=96$ \AA\hs when considering a 'standard' conversion between Ly$\alpha$ luminosity, UV luminosity density \citep{Kennicutt98}, which are L$_{\nu,{\rm UV}}=$SFR$\times 8.0\times 10^{27}$ erg s$^{-1}$ Hz$^{-1}$ and $L_{\rm{Ly\alpha}}=$SFR$\times 1.0\times 10^{42}$ erg s$^{-1}$. We emphasize that an implicit assumption in the conversion is a Salpeter IMF in the mass range 0.1-100 \Msun for solar abundance. In our numerical experiments, EW$_{0}=110\pm 12$\hs\AA, which is a result of the lower metallicities and the higher upper limit in the IMF mass range assumed in SLUG (details are given in \S\ref{sec:slug}). These minor systematic corrections due to different conditions in stellar metallicity, IMF or star formation history won't affect our main conclusions. Throughout this paper, we will adopt  EW$_{0}=110$\hs\AA.

\subsection{Stochasticity}

The description of star formation is probabilistic. At its core we
find the concept of the Initial Mass Function (IMF), $\phi(m)$, which
describes the relative abundance of stars of a given mass to be
formed. The IMF can be mostly determined by two pieces of
information. The first one is the mass
interval, $m_{\rm min}-m_{\rm max}$ that the newly formed stars can
have. The second is the change of relative abundance for different stellar
masses, which is usually parametrized as $\phi(m)\propto m^{-\gamma}$,
if $\gamma>0$ less massive stars are more probable to
form that high mass stars.   

For large masses of newly formed stars it is possible to guarantee a
good sampling of the IMF. For a low rate of star formation the poor
sampling of the IMF makes possible that the rarer most massive stars
near $m_{\rm max}$ don't appear in a burst of star formation. 

Star formation is also thought to proceed in clusters. This provides
another important astrophysical element in the statistical description
of star formation that can be impacted by stochasticity effects. In
analogy to the stellar mass function, one can also parametrize the
abundance of stellar clusters by the cluster mass function (CMF). Low
values of the star formation rate can also affect the sampling of 
the CMF. 

Furthermore, stars of different mass have different evolutionary
tracks. The inhomogeneous IMF sampling will produce fluctuating
stellar populations with different luminosity histories, this can
impact the ratio of ionizing to far UV photons produced as a function
of time.

The effect of stochasticity has been explored before in the
production of H$\alpha$ in dwarf galaxies in the local universe \citep{2011ApJ...741L..26F,2010A&A...512A..79H}, but not on the effect on the EW of Lyman$\alpha$ emitting galaxies. Our aim is to quantify this impact and estimate how current and future
observational campaigns can be affected by these effects.  

\subsection{Grids of SLUG models}
\label{sec:slug}

We used {\sc SLUG} (`Stochastically Light Up Galaxies') a fully stochastic code for synthetic photometry
of galaxies \citep{2011ApJ...741L..26F,dasilva12}. The code
samples the CMF and IMF of a star forming galaxy by forming individual star
clusters and following their spectral evolution. At low values of a
time averaged continuous SFR the series of bursts associated with the
cluster formation start to be visible. Furthermore, low SFR values 
imply a low number of stars to sample the IMF, lowering the probability of finding massive stars. All the
effects of finite sampling in mass and time have an impact on the
instantaneous strength of the Lyman$\alpha$ emission line. {\sc SLUG}
assumes that a fraction $f_{c}$ of the total stellar population is
formed in clusters. In that case, the clusters also sample a cluster
mass function(CMF) $\psi(M_{\rm ecl})=M_{\rm ecl}^{-\beta}$. The stars
in the cluster are then populated according to the selected IMF.

In this stochastic framework, a galaxy with a time averaged constant
SFR does not level to a constant SED, but instead it continuously
fluctuates in time. For a sample of independent galaxies with the same
average SFR one could thus associate a distribution of line and
continuum intensities.  

All models used in this paper are based on a Salpeter IMF in the mass
intervals $0.08$--$120$ \Msun. We consider two kinds of models of varying
metallicity $Z=4.0\times 10^{-4}, 4.0\times 10^{-3}$ (which corresponds to $Z=0.02 Z_{\odot}$, $Z=0.2Z_{\odot}$).  We assume that
all the stars are formed in clusters, $f_{c}=1$. Assuming unclustered star formation, $f_{c}=0$, reduces the scatter in the flux fluctuations we report here by about an order of magnitude \citep{dasilva12}. The CMF is described
by the cluster mass interval is $20-10^7$\Msun\hs and the slope is
$\beta=2.0$. The stellar libraries are Padova asymptotic giant branch
(AGB) tracks. The stellar Spectral Energy Distributions (SED) correspond to \cite{Lejeune1,Lejeune2} where the Wolf-Rayet stars treatment follows \cite{Smith02} and \cite{Hillier98}. We run 500 galaxies for each metallicity and star
formation rate comprised in the range $10^{-3}$--$1$ \Msun/yr.  Table
1 lists the exact values for the SFR and metallicity that were
used. Each model runs for a period of 200 Myr with outputs every 1 Myr. Running for a period longer than 100 Myr ensures a stable base level for the mean FUV and ionizing fluxes.

We use the last 10 outputs per galaxy to slightly increase the
statistical sample. Being pessimistic, one can assume that each simulated system contributes only 1 independent datum instead of 10, from that point of view, one can be sure that all the results for the PDF and derived quantities are accurate within $1/500\sim 0.02$. In other words, one might be concerned that the 10 outputs of the same galaxy may not be independent and might bias the PDF. To quantify this we perform the following test: for 4000 galaxies with a star formation rate $8.0\times 10^{-3}$ \Msun yr$^{-1}$ we calculate the PDF in two different ways. First by taking 10 outputs of 400 galaxies and second by using only 1 output for each one of the 4000 galaxies. We find that the absolute difference between the PDF($\mathcal{M}$) constructed from the two different samples is always on  the order of $\sim 0.02$ and conclude that a possible bias is negligible. We perform an additional test by taking 1 output and 2 outputs for 4000 galaxies, with spectral outputs spaced by $100$Myr during $500$Myr and find that the difference between the PDF($\mathcal{M}$) constructed from these samples is always on the order of $\sim 0.05$ and conclude that a possible bias from the chosen timesteps is also negligible.

For each galaxy we obtain the flux per unit frequency 
$F_{\nu,{\rm FUV}}$ at a wavelength of 1540 \AA, together with the ionizing photon rate $Q_{H}$.  These quantities are used to calculate the EW as described
by Eq (\ref{eq:def}). The value for EW$_{0}$ is calculated as the
arithmetic average EW for 30 models of galaxies with a SFR of $5$ \Msun yr$^{-1}$, we obtain EW$_{0}=110 \pm 12$ where the error quoted is estimated from the standard deviation from the mean.

\section{Results}
\label{sec:results}
\begin{table*}
\caption{Best fit values for the parameters used in the conditional PDF $P(\mathcal{M}|{\rm SFR})$ (see Eq.\ref{eq:PDF-EW}).}.
\label{table:fit}

\begin{tabular}{cccccc} \hline\hline
  SFR (\Msun yr$^{-1}$)& $Z/Z_{\odot}$ & ${\mathcal M}_0$& $\alpha$ &
  $\gamma$ & $P_{0}$ \\\hline
0.0010	&0.02	&0.75$\pm$0.03	&0.13$\pm$0.03	&2.67$\pm$0.10	&1.15$\pm$0.05\\
0.0015	&0.02	&0.71$\pm$0.02	&0.44$\pm$0.03	&2.73$\pm$0.09	&1.44$\pm$0.04\\
0.0025	&0.02	&0.71$\pm$0.03	&0.74$\pm$0.06	&2.82$\pm$0.16	&1.66$\pm$0.07\\
0.0040	&0.02	&0.76$\pm$0.01	&1.03$\pm$0.03	&3.49$\pm$0.09	&1.86$\pm$0.03\\
0.0063	&0.02	&0.80$\pm$0.01	&1.21$\pm$0.02	&3.83$\pm$0.05	&1.97$\pm$0.01\\
0.010	&0.02	&0.77$\pm$0.01	&1.81$\pm$0.10	&4.01$\pm$0.10	&2.37$\pm$0.06\\
0.015	&0.02	&0.75$\pm$0.01	&2.15$\pm$0.08	&3.95$\pm$0.11	&2.57$\pm$0.03\\
0.025	&0.02	&0.75$\pm$0.01	&2.79$\pm$0.08	&4.12$\pm$0.09	&2.89$\pm$0.03\\
0.040	&0.02	&0.79$\pm$0.01	&3.18$\pm$0.06	&4.66$\pm$0.06	&3.09$\pm$0.02\\
0.063	&0.02	&0.76$\pm$0.01	&4.31$\pm$0.07	&4.52$\pm$0.06	&3.43$\pm$0.01\\
0.1	&0.02	&0.77$\pm$0.01	&5.12$\pm$0.13	&4.57$\pm$0.10	&3.64$\pm$0.03\\
1.0	&0.2	&0.85$\pm$0.01	&12.49$\pm$0.36	&6.01$\pm$0.11	&5.23$\pm$0.05\\
0.0010	&0.2	&0.64$\pm$0.05	&0.013$\pm$0.040	&2.25$\pm$0.15	&1.20$\pm$0.10\\
0.0015	&0.2	&0.58$\pm$0.04	&0.39$\pm$0.05	&2.38$\pm$0.14	&1.62$\pm$0.09\\
0.0025	&0.2	&0.64$\pm$0.02	&0.62$\pm$0.04	&2.73$\pm$0.12	&1.72$\pm$0.06\\
0.0040	&0.2	&0.78$\pm$0.02	&0.70$\pm$0.05	&3.60$\pm$0.18	&1.74$\pm$0.06\\
0.0063	&0.2	&0.71$\pm$0.01	&0.70$\pm$0.05	&3.62$\pm$0.14	&2.10$\pm$0.05\\
0.010	&0.2	&0.69$\pm$0.01	&1.52$\pm$0.07	&3.96$\pm$0.15	&2.50$\pm$0.05\\
0.015	&0.2	&0.72$\pm$0.01	&1.85$\pm$0.09	&4.34$\pm$0.18	&2.61$\pm$0.06\\
0.025	&0.2	&0.64$\pm$0.01	&3.08$\pm$0.11	&3.90$\pm$0.09	&3.30$\pm$0.03\\
0.040	&0.2	&0.70$\pm$0.01	&2.87$\pm$0.13	&4.50$\pm$0.17	&3.26$\pm$0.05\\
0.063	&0.2	&0.71$\pm$0.01	&3.46$\pm$0.16	&4.87$\pm$0.19	&3.58$\pm$0.05\\
0.1	&0.2	&0.70$\pm$0.01	&4.42$\pm$0.13	&4.69$\pm$0.10	&3.93$\pm$0.03\\
1.0	&0.2	&0.76$\pm$0.01	&11.52$\pm$0.43	&6.08$\pm$0.15	&5.87$\pm$0.07\\\hline\hline
\end{tabular}  
\end{table*}

\begin{figure}
\includegraphics[width=0.50\textwidth]{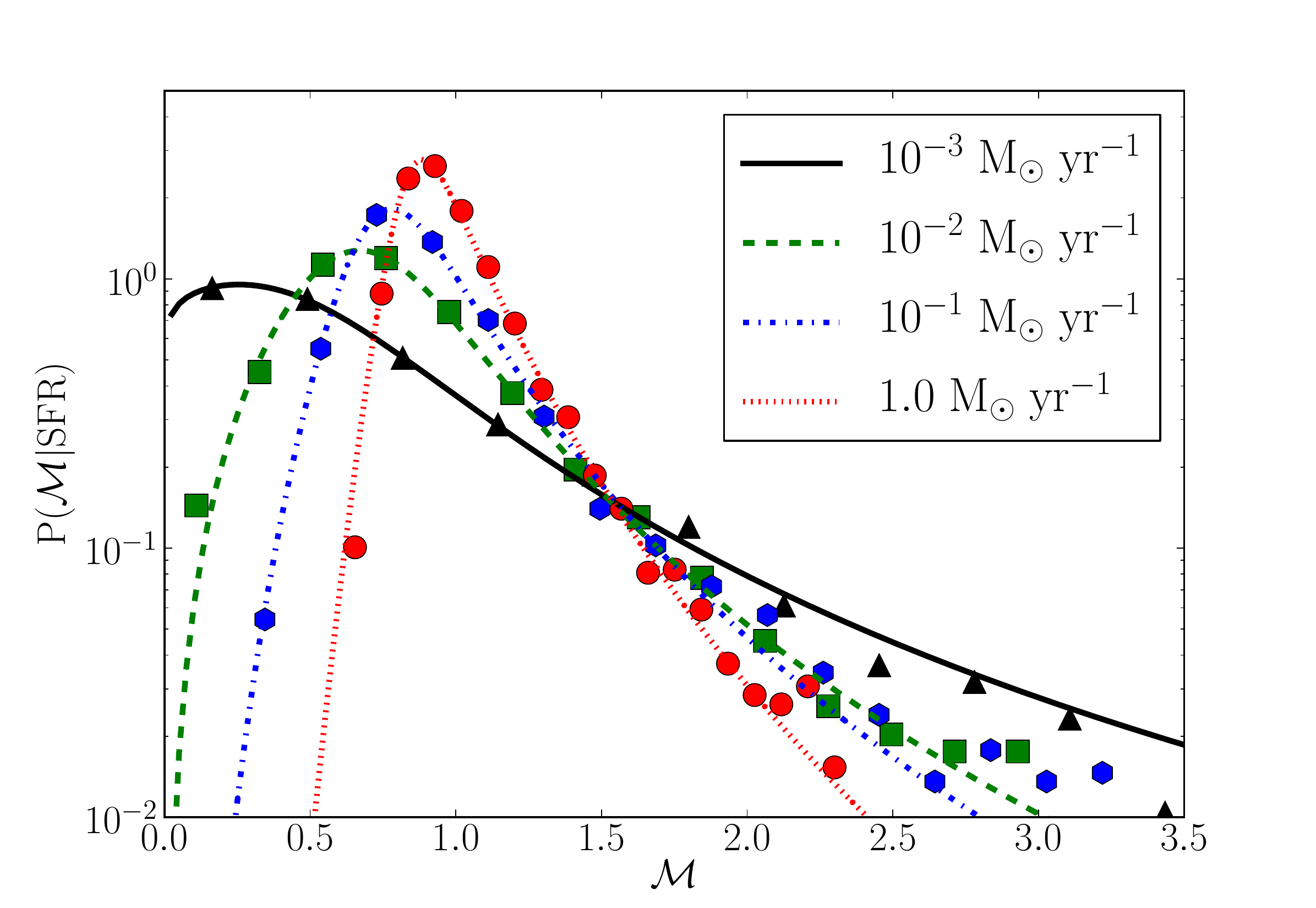}
\caption{Different Probability Density Distributions for ${\mathcal M}$,
  the normalized EW. Each symbol corresponds to the binned SLUG results for a different  SFR=$10^{-3}$, $10^{-2}$, $10^{-1}$ and $1$\Msun yr$^{-1}$ and a metalliciticy f $0.20 Z_{\odot}$. The lines are the best fit of Eq. (\ref{eq:PDF-EW}) to the points.}
\label{fig:PDF-EW}
\end{figure}

The net effects of stochasticity introduce a SFR dependent scatter around a
central value. We describe this effect as a Probability Density Distribution
(PDF) for the ${\mathcal M}$ variable. We parametrize the PDF by a
double power law in order to better quantify how it depends from
changes in the SFR and the metallicity, 

\begin{equation}
P({\mathcal M}|{\rm SFR}) = P_{0} \left[ \left(\frac{{\mathcal M}}{{\mathcal M}_{0}}\right)^{-\alpha} + \left(\frac{{\mathcal M}}{{\mathcal M}_{0}}\right)^{\gamma}\right]^{-1}.
\label{eq:PDF-EW}
\end{equation}

Figure \ref{fig:PDF-EW} presents the results for the PDF
$P({\mathcal M}|{\rm SFR})$, the symbols represent the results from the simulations and the lines the fit to those points. Given the limited number of models the PDF is reliable only for values $P({\mathcal M}|{\rm SFR})>10^{-2}$, likewise the deviation of the fit with respect to the simulation points is on the order of $1 \times 10^{-2}$. The star formation rate enters through the constants $\mathcal{M}_0$, $\alpha$, $\gamma$, and $P_0$.
 The expected result for a perfectly sampled IMF is a delta function around ${\mathrm{M}}=1$ \citep{1999ApJS..123....3L}, while our results show a wide distribution. Each line in
Figure \ref{fig:PDF-EW} corresponds to a different SFR$=1\times 10^{-3}$, $10^{-2}$, $10^{-1}$ and $1.0$
in units of \Msun yr$^{-1}$ for a metallicity of $0.02 Z_{\odot}$.
The best fit values together with their uncertainty from the fitting procedure for all the SLUG
models are listed in Table 1. In these values we get a precise
quantification on how the PDF shape changes with the SFR. From these results we 
confirm that the effect of the metallicity is a second order effect
shaping the EW PDF when compared to the influence of the SFR. 

For high values of the SFR$>10^{-1}$ \Msun yr$^{-1}$ the low end,
${\mathcal M}<1$ of $P({\mathcal M})$ increases sharply compared to
the slower decrease in the high end ${\mathcal M}>1$. In this regime the peak
of the distribution is around the expected value of ${\mathcal M}=1$
for a perfectly sampled IMF. 

This quantitative behaviour presents consequences for the
interpretation of high EW systems. On the upper end, on can achieve
values close to what is expected to be limit of star forming galaxies:
$240$ \AA\hs  which corresponds to ${\mathcal M}=2.2$ in our models. For SFR$<10^{-2}$ \Msun yr$^{-1}$, the low end of the PDF shows a bulkier shape that confirms the existence of systems with ${\mathcal M}<1$ purely as a consequence of their intrinsic star formation stochasticity. We quantify and elaborate in the next section the implications of these results.

\section{Discussion}
\label{sec:discussion}

\subsection{High EW outliers and apparent Lyman$\alpha$ escape fraction}
\begin{figure}
\includegraphics[width=0.50\textwidth]{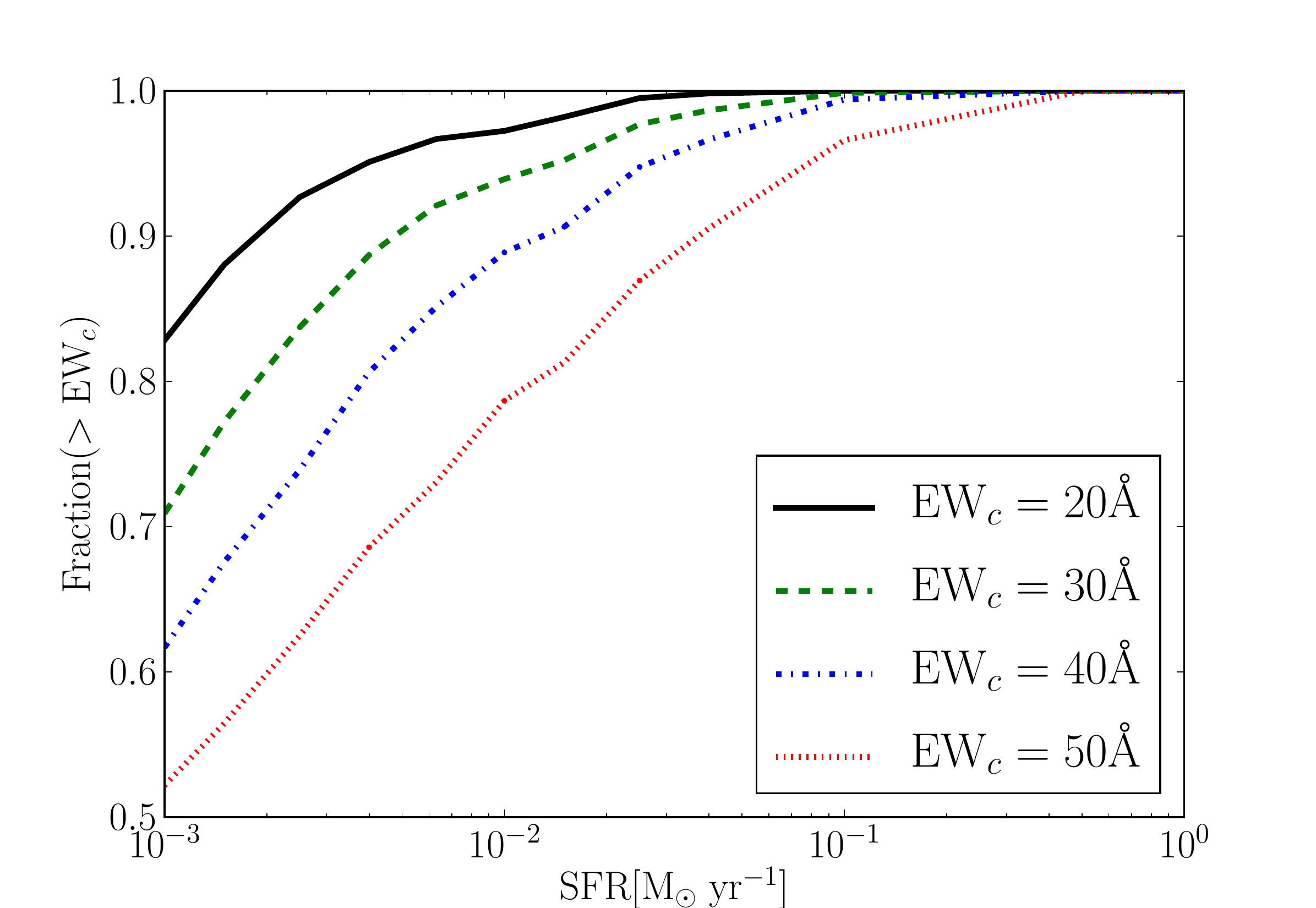}
\caption{Fraction of galaxies as a function of
  SFR that have an EW larger than a fixed value EW$_{c}$. Each line represents different threshold in EW$_{c}=20$, $30$, $40$ and $50$\AA\hs. This figure quantifies the impact   of the stochasticity on the fraction of confirmed LAEs based on a  minimal EW detection threshold.} \label{fig:EW_cut}
\end{figure}

The ensuing discussion will focus on the interpretation of high and low EW systems that naturally emerge through stochasticity. High EW vales might lead one to propose the presence of very low metallicity (or population III) stars while low EW values could cause some galaxies not to be selected as a LAE in very deep narrowband surveys.

A common assumption in the search for galaxies with Population III stars is the following: there is an upper EW value to consider a star forming galaxy to be normal. This threshold value is taken to be EW$_{\rm th}\sim 240$ \AA\hs but, as we show in the previous section, galaxies with similar or larger EW values can be expected by stochasticity effects without demanding Pop III stars.  The threshold value EW$_{\rm th}=240$ \AA \hs corresponds to ${\mathcal M}=2.2$ (Eq. \ref{eq:M}). We can then quantify the fraction of such galaxies that have EW that exceed the threshold by using Eq. (\ref{eq:PDF-EW}). For galaxies with star formation rates in the range $0.1$--$1$ \Msun yr$^{-1}$ we find that there would be between $0.5-2.0\%$ of systems with EW values larger than EW$_{\rm th}$. We calculate same fraction directly from the $500\times 10$ points used to build the PDF and find that between $1\%$ to $3\%$ among them have an EW larger than $EW_{\rm th}$.

This suggests that a minimal threshold in EW is not enough to assert that an individual galaxy presents signatures of population III stars. However, also the effects of stochasticity cannot be observationally evaluated from the measurement of an individual galaxy. Only statistical studies of galaxy samples are an appropriate tool to find clear signposts either of special stellar populations or stochasticity.

Narrowband surveys for LAEs often employ color cuts that can be approximated by requiring that the EW of a galaxy exceeds threshold value, which typically lies in the range EW$_{\rm cut}=20-65$ \AA\hs (see \S~\ref{sec:introduction}). The result of previous sections show that  low values of the SFR$<10^{-2}$ M$_{\odot}$ yr$^{-1}$ the EW can be lower than half of the expected EW$_{0}$. In a survey that adopts color-cuts that correspond to a high EW$_{\rm cut}$, one could simply miss those objects due to their low intrinsic EW. This shows how stochasticity can induce selection bias in LAEs observations. 

Not surprisingly, the impact of stochasticity is highly SFR dependent. The strong dependence of the shape of the $\mathcal{M}$--PDF on SFR causes a larger fraction of systems to be found below a fixed $\mathcal{M}$ threshold for detection as the SFR decreases. In Figure \ref{fig:EW_cut} we show the fraction of LAEs that are missed as a function of the star formation rate, for different cuts in the EW expressed by EW$_{c}$. These fractions are calculated directly from the SLUG models, providing an uncertainty on the order of $\sim 0.01$ in the reported fraction of detected galaxies, the discrepancy of calculating the same fractions from the fitting function is in all cases smaller than $\sim 0.02$. The value of EW$_{c}=20$ \AA\hs corresponds to a common cut used in present day observations \citep[e.g.][]{Ouchi08}. A cut of EW$_{c}=50$ \AA\hs represents the cut used in the early observational attempts to find Lyman $\alpha$ emitting galaxies \citep[e.g.][]{Rhoads00}. In \S~\ref{sec:LFEW} we investigate how such a bias may affect predicted number densities of LAEs in simplified (but often employed) models of LAEs.

\subsection{The connection between the LAE-LF and the UV-LF}
\label{sec:LFEW}
In this section we study the potential impact of stochasticity on the predicted Ly$\alpha$ luminosity functions \& equivalent width distribution of narrowband selected galaxies (i.e. LAEs). We compare two very simple models to isolate and highlight the impact of stochasticity. In the first model we ignore stochastic ffects. We further assume that each star forming galaxy has a Ly$\alpha$ emission line with EW$=$EW$_0=110$ \AA\hs (see \S~\ref{sec:ew}). In the second model, we instead assume that the Ly$\alpha$ emission line of a star forming galaxy is drawn from a PDF which is set entirely by stochasticity effects. We therefore assume that the function $P({\rm EW}|{\rm SFR})$ is given by Eq~\ref{eq:PDF-EW} for the second model. Finally, we further assume that there is a one-to-one correspondence between SFR and $M_{\rm UV}$ (given by the standard conversion factor between SFR and UV continuum flux, see \S~\ref{sec:ew}), and therefore\footnote{ Of course, in reality stochasticity introduces a dispersion in the value of SFR at a given $M_{\rm UV}$, and the proper EW-PDF for a fixed $M_{\rm UV}$ is $P({\rm EW}|M_{\rm UV})=\int d\hs {\rm SFR}\hs  P({\rm EW}|{\rm SFR})P({\rm SFR}|M_{\rm UV})$. Because the PDF $P({\rm EW}|{\rm SFR})$ varies relatively weakly with SFR, this proper calculation barely changes our results at all.} that $P({\rm EW}|{\rm SFR})\equiv$ $P({\rm EW}|M_{\rm UV})$. Under these assumptions, we can predict the LAE Ly$\alpha$ luminosity function, $\Psi(L_{\alpha})$, for both models as \citep{DW12}

\begin{eqnarray}
\Psi(L_{\alpha})={\rm ln} 10 \int dM_{\rm UV}\hs \phi(M_{\rm UV})\times {\rm EW_{\rm c}} \times P({\rm EW_{\rm c}}|M_{\rm UV}),
\end{eqnarray} where $\Psi(L_{\alpha})d \log L_{\alpha}$ denotes the comoving number density of LAEs in the luminosity range $\log L_{\alpha} \pm d\log L_{\alpha}/2$. Furthermore, $\phi(M_{\rm UV})$ denotes the comoving number density of star forming galaxies with absolute UV-magnitude in the range $M_{\rm UV} \pm dM_{\rm UV}/2$. The equivalent width EW$_{\rm c}$ denotes the EW which corresponds to a Ly$\alpha$ line of luminosity $L_{\alpha}$ for an absolute UV magnitude $M_{\rm UV}$ (see Dijkstra \& Wyithe 2012 for details). We focus our analysis on $z=5.7$ and compare to the observed LAE Ly$\alpha$ luminosity functions at this redshift by \citet{Ouchi08}. We take the observed $\phi(M_{\rm UV})$ at $z \sim 6$ from \citet{2007ApJ...670..928B} and `rescale' it to $z=5.7$ assuming that $M^*_{\rm UV}$ evolves with redshift as $M^*_{\rm UV}=-21.02+0.36(z-3.8)$ \citep{bouwens08}.

\begin{figure}
\label{fig:lf}
\includegraphics[width=0.50\textwidth]{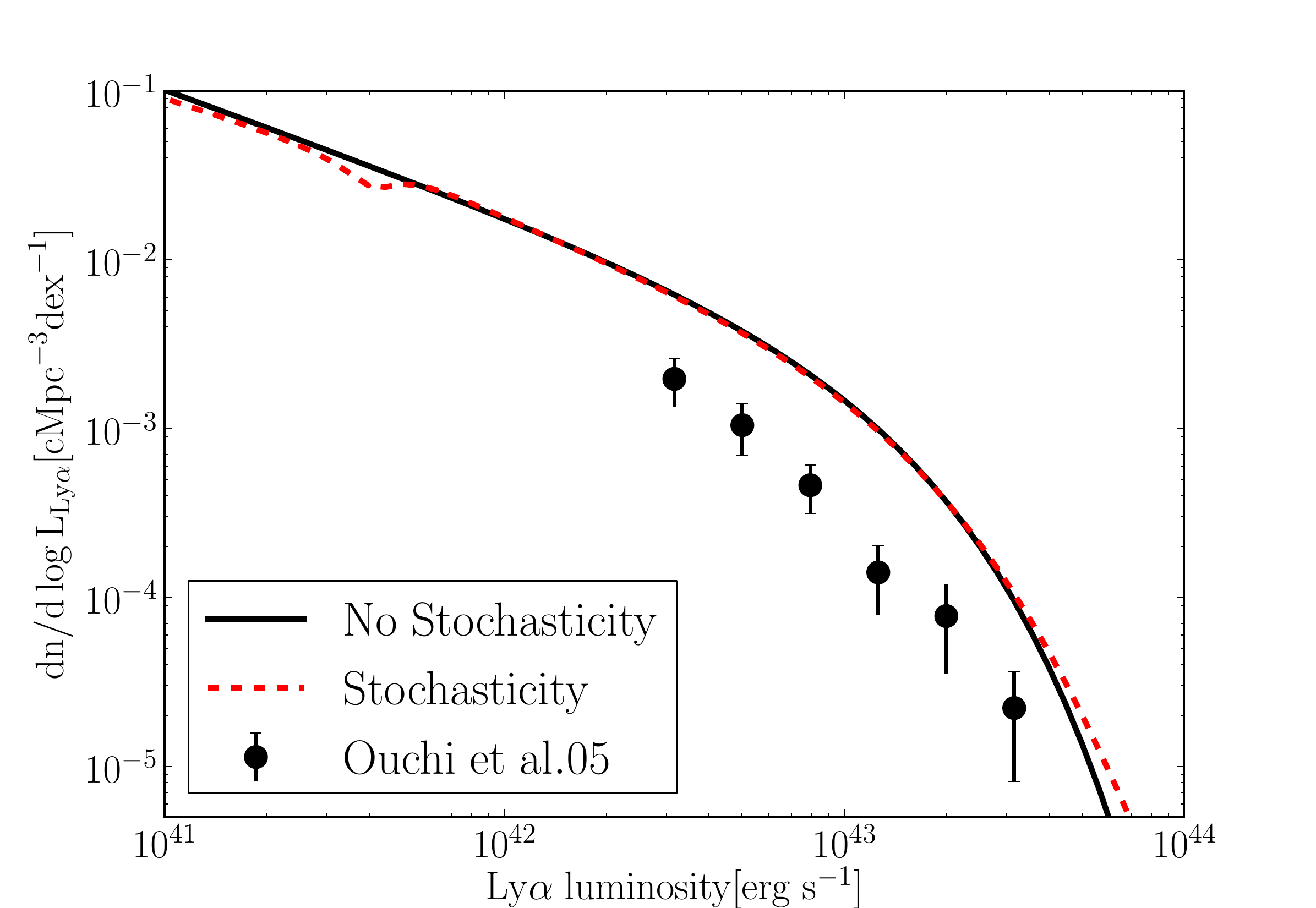}
\caption{This Figure shows predicted Ly$\alpha$ luminosity function of LAEs for simplified models with (without) the effects of stochasticity included as the {\it red dashed} ({\it black solid line}). Also shown are the observations taken from Ouchi et al. (2008, {\it black filled circles}). The main point of this plot is that stochasticity barely affects the predicted luminosity functions. Both models predict more LAEs than is observed, as we have not attempted to correct for dust extinction and/or intergalactic scattering, which would both reduce the predicted luminosity functions.} 
\end{figure}

Figure~\ref{fig:lf} shows the luminosity function for the first model (no stochasticity) as the {\it black solid line}, and the second model (with stochasticity) as the {\it red dashed line}. The data points represent observations by \citet{Ouchi08}.  Both models are remarkably similar. The reason for this similarity is non-trivial: stochasticity decreases the `most likely' EW for star forming galaxies compared to models that do not include this effect (see Fig~\ref{fig:PDF-EW}). This decrease alone would reduce the predicted Ly$\alpha$ luminosity functions. However, it is compensated for by the fact the $\mathcal{M}$-PDF has a tail to values $\mathcal{M}>1$. The presence of this tail allows for an additional contribution to $\Psi(L_{\alpha})$ (at a given $L_{\alpha}$) by UV-faint galaxies that would not be able to contribute in the absence of stochasticity. Our prediction further assumes that $P({\rm EW}|{\rm SFR}> 1 M_{\odot}\hs{\rm yr}^{-1})$=$P({\rm EW}|{\rm SFR=1}M_{\odot}\hs{\rm yr}^{-1})$, and therefore likely overestimates the impact of stochasticity at SFR$> 1 M_{\odot}$ yr$^{-1}$ (which corresponds to $L_{\alpha}> 10^{42}$ erg s$^{-1}$). We therefore safely conclude that stochasticity barely impacts predictions of Ly$\alpha$ luminosity functions. This result is not sensitive to the uncertainties in the fitting functions for the PDF(${\mathcal{M}|{\rm SFR}}$) used here. Also shown for completeness is the observed $z=5.7$ LAE luminosity function from Ouchi et al. (2008). Both models clearly predict more LAEs than is observed. This is not surprising as we have not attempted to correct for dust extinction and/or intergalactic scattering. Both these effects likely reduce the predicted Ly$\alpha$ luminosity at a fixed number density of objects, which would bring down the predicted luminosity functions.

The EW-PDF for a sample of Ly$\alpha$-selected galaxies is given by (Dijkstra \& Wyithe 2012)
\begin{eqnarray}
\nonumber
P({\rm EW}|{\log L_{\alpha,{\rm min}}},{\log L_{\alpha,{\rm max}}})=\\ 
\hspace{0mm}{\mathcal N}\int_{\log L_{\alpha,{\rm min}}}^{\log L_{\alpha,{\rm max}}} P({\rm EW}|M_{\rm UV,c})\phi(M_{\rm UV,c})d\log_{10} L_{\alpha},
\end{eqnarray} where $
L_{\alpha,{\rm min}}$ ($L_{\alpha,{\rm max}}$) denotes the minimum (maximum) Ly$\alpha$ luminosity of the galaxies in the sample. Furthermore, $M_{\rm UV,c}$ denotes the UV magnitude for which a Ly$\alpha$ line of luminosity $L_{\alpha}$ corresponds to an equivalent width EW (again see Dijkstra \& Wyithe 2012 for details). We focus on the impact of stochasticity on the predicted PDF for $\log L_{\alpha,{\rm min}}=42$ and $\log L_{\alpha,{\rm max}}=43$ ({\it blue dotted line} (which corresponds approximately to $1\hs M_{\odot}\hs{\rm yr}^{-1}<$SFR$< 10\hs M_{\odot}\hs{\rm yr}^{-1}$, and roughly to the range of observed Ly$\alpha$ luminosities). For comparison, we also show a case with $\log L_{\alpha,{\rm min}}=40$ and $\log L_{\alpha,{\rm max}}=42$ ({\it red dashed line}, which corresponds approximately to $0.01\hs M_{\odot}\hs{\rm yr}^{-1}<$SFR$< 1\hs M_{\odot}\hs{\rm yr}^{-1}$). The choice $\log L_{\alpha,{\rm min}}=40$ is a bit arbitrary, but was motivated by the notion that galaxies with SFR$\sim 10^{-2}\hs M_{\odot}$ yr$^{-1}$ can contribute significantly to the overall ionizing photon budget during the Epoch of Reionization (see Kuhlen \& Faucher-Giguere 2012). 

\begin{figure}
\label{fig:ewpdf}
\includegraphics[width=0.50\textwidth]{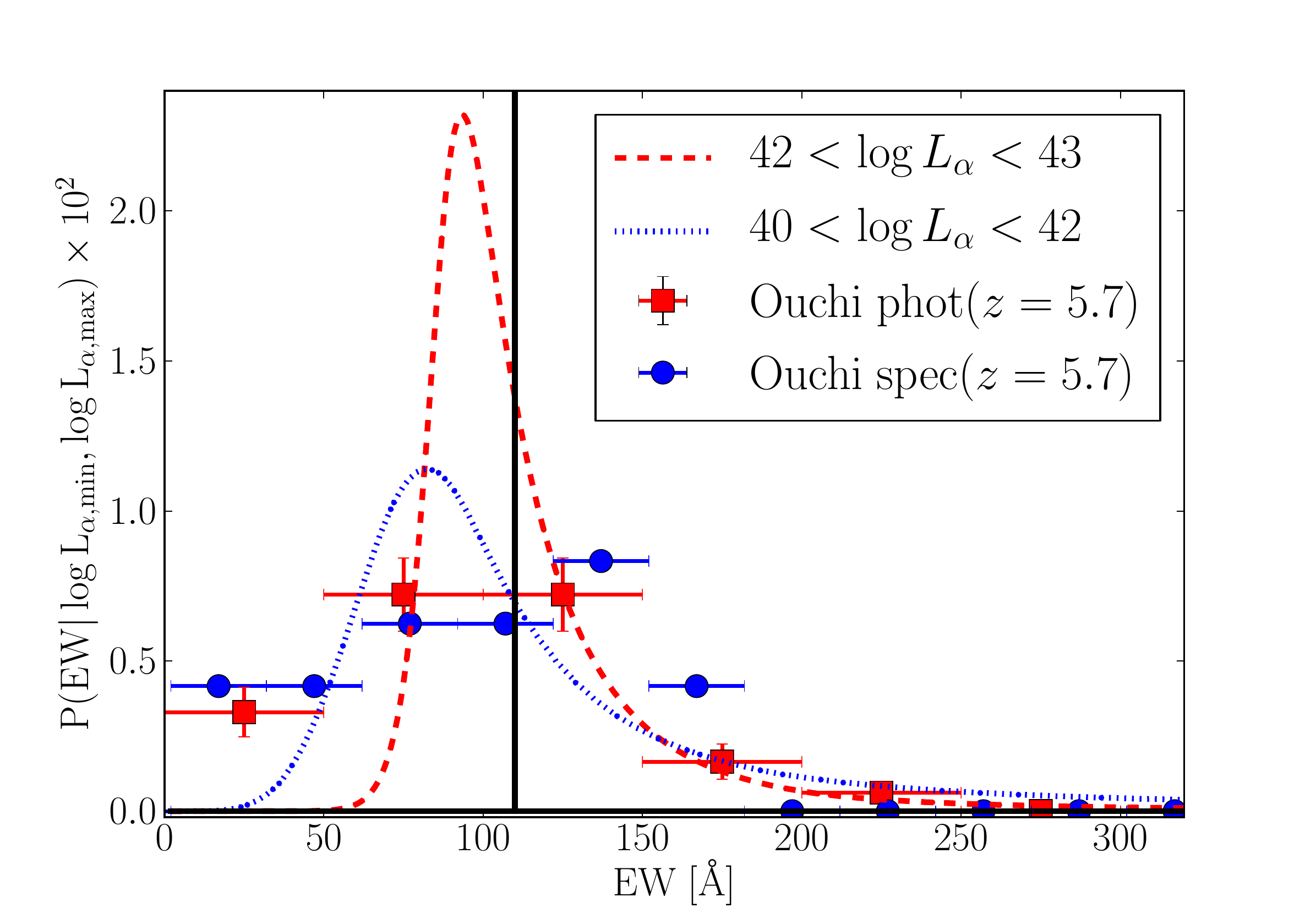}
\caption{This Figure shows the predicted EW-PDF for a sample of Ly$\alpha$ selected galaxies. The {\it solid black line} shows the Dirac delta function for the model with no stochasticity effects. The {\it red dashed line} ({\it blue dotted line}) shows the predicted EW-PDF for models which include stochasticity effects for Ly$\alpha$ selected galaxies with $L_{\alpha}=10^{40}-10^{42}$ erg s$^{-1}$ ($L_{\alpha}=10^{42}-10^{43}$ erg s$^{-1}$, see text). Also shown for comparison is the EW-PDF observed by Ouchi et al. (2008; the {\it red filled circles} represent their photometric data, while the {\it blue filled circles} represent their spectroscopic sample). This Figure illustrates that stochasticity may help explain the large observed values of EW of LAEs at z=5.7, although it fails to fully explain the observed EW-PDF.}
\end{figure}

The {\it black solid line} in Figure~\ref{fig:ewpdf} shows the predicted EW-PDF for the model with no stochastic effects (which is simply a delta-function). The predicted PDF for the models that include stochasticity peak at the lower values of EW, and have tails that extend to values that greatly exceed EW$=100$ \AA. Also shown is the observed EW-PDF for a sample of photometrically  selected  ({\it red filled squares}) and spectroscopically confirmed ({\it blue filled circles}, taken from Ouchi et al. 2008) LAEs. Clearly, stochastic effects may help to explain large observed values of EW. However, additional processes such as dust extinction and/or intergalactic scattering, which impacts differently the UV continuum and the Ly$\alpha$ line \citep{Forero11}, are needed to fully explain the observed EW-PDF of LAEs at z=5.7.

\subsection{Stochastic effects on the Lyman continuum}

The distribution of EW at a given averaged SFR is the result of fluctuations in the fluxes of ionizing and far UV continuum (non ionizing) photons. The ionizing photon production is dominated by short--lived O \& B stars, while less massive A stars can contribute significantly to the non-ionizing photon flux. Because of the different lifetimes of these stars of different masses, the ionizing and non ionizing fluxes -- and therefore their ratio -- vary on different timescales and with different amplitudes.

Simultaneous fluctuations in the ionizing and non-ionizing fluxes can thus impact observational estimates of the ionizing escape fraction, which is usually estimated from a \textit{relative escape fraction}:

\begin{equation}
f_{{\rm esc, rel}} = \frac{(f_{\rm LyC}/f_{1500})_{\rm obs}}{(f_{\rm LyC}/f_{1500})_{\rm stel}} \exp{(\tau_{\rm IGM})},
\end{equation}

where the $(f_{\rm LyC}/f_{1500})_{\rm stel}$ is the intrinsic flux ratio for a stellar population model, $(f_{\rm LyC}/f_{1500})_{\rm obs}$ is the observed flux ratio and $\tau_{\rm IGM}$ is the LyC optical depth of the IGM at the redshift of emission.

Based on Subaru/Suprime-Cam imaging \citet{2009ApJ...692.1287I} claimed a
detection of 10 LAEs and 3 LBGs at $z\sim 3.1$ with a significant
emission in LyC  out of a sample of 125 LAEs and 73 LBGs. Recent measurements seem to confirm such measurement for 8 LAEs
using spectroscopy with VLT/VIMOS and SUBARU/FOCAS. The results
indicate a LyC excess with respect to non-ionizing UV at $\sim 1500$\AA\hs
 expected by standard stellar populations
 \citep{2011MNRAS.411.2336I}. However, observation targeting LyC in
 LBGs at $z\sim 1.3$ using deep Hubble Space Telescope imaging do not
 find any excess \citep{2010ApJ...723..241S}.

The sources present an excess of a factor of $\sim 2-3$ times larger
than the expected intrinsic stellar ratio for
$f_{{\rm LyC}}/f_{1500}$. Some theoretical analysis requires a very top
heavy IMF to explain the observed ratios or fractions of $10\%$ in
mass of very young $<1$ Myr of extremely metal poor $Z<1.0\times
10^{-4} $Z$_{\odot}$ or metal free stellar populations
\citep{2010MNRAS.401.1325I,2011MNRAS.411.2336I}

The stochasticity effects present an important element to be considered in such
analysis: The ratio ${\mathcal M}$ can
be directly interpreted as the excess of the $f_{LyC}/f_{1500}$ ratio
with respect to the constant expected value. The star
formation rates for the objects with LyC excess are on the order of
$\sim 1$ \Msun \hs yr$^{-1}$ ($M_{\rm UV} \gsim -19$, see Fig~5 of Siana et al. 2010 and where we ignore dust in the conversion from $M_{\rm UV}$ to SFR). {In this range $10\pm 1 \%$ of the galaxies will show an excess $1.5-2.0$ times the expected stellar $f_{LyC}/f_{1500}$ ratio only due to stochasticity (Figure 1),
this is consistent with the fraction of galaxies observed with those
characteristics.}
                                
The stochasticity might also explain why the LyC excess is not
observed in the recent $z\sim 1.3$ Hubble imaging by \citet{2010ApJ...723..241S}. The LBGs observed at that
redshift have systematically restframe magnitudes $M_{\rm UV}<-20$,
while the LyC excess sources are fainter than this limit. This
transition roughly corresponds to a SFR $\sim 5$ \Msun \hs yr$^{-1}$
assuming a standard conversion factor, while the fainter systems
reported with an excess have $M_{\rm UV}\sim -18$ and SFR $\sim
1$ \Msun \hs yr$^{-1}$. The dependence of the stochasticity effects on
the SFR make it more likely to detect the LyC excess in the fainter
sample. Probably the stronger effect is of statistic origin, given
that the observations at $z\sim 1.3$ target 15 LBGs, while at $z\sim
3.0$ a set of 198 galaxies were observed, this provides one order of
magnitude less objects to sample the $P({\mathcal M})$ distribution and
find galaxies with a LyC excess.  

\section{Conclusions}
\label{sec:conclusions}

In this paper we have investigated the quantitative impact of stochastic sampling of the stellar initial mass function (IMF)
 on the Ly$\alpha$, and ionizing photon emissivity of dwarf galaxies.  We have used the SLUG code to simulate the spectral energy distribution for restframe wavelengths $\lambda \lsim 2000$ \AA\hs of galaxies with time-averaged star formation rates in the range $(10^{-3}-1)$ \Msun \hs yr$^{-1}$. 
 
 From our predicted Far UV continuum flux density (at $\langle \lambda \rangle =1500$ \AA) and the ionizing flux density (at $\langle \lambda \rangle <912$ \AA) we construct probability distribution function (PDF) for the variable ${\mathcal M}$ that represents the ratio of the measured restframe equivalent width (EW) of the Ly$\alpha$ line to the
expected constant value in the absence of stochasticity (EW$_0$), i.e. $\mathcal{M}\equiv$EW/EW$_0$.

We find that the $\mathcal{M}$--PDF can be represented by a double power law in ${\mathcal M}$, whose parameters are highly dependent on the SFR. We emphasize that the results derived from these experiments are sensitive to the degree of the clustering properties of the stars. In the case of $f_{c}=0$ where the star formation is completely unclustered, the dispersion around the mean is reduced by a factor of 10 \citep{dasilva12}. The results are also sensitive to the star formation history assumed: in the case of a bursty star formation history, one could expect the scatter around the mean for $\mathcal{M}$ to be higher than what we derive here.

Our results show that it is possible for galaxies to have both extremely low and high ${\mathcal M}$ 
values -- especially at increasingly low SFR -- and investigate implications of this result. In particular, we show how the existence of galaxies with low ${\mathcal M}$
values can induce an observational bias if galaxies are selected by cuts in
the equivalent width (as in narrowband surveys for LAEs). This bias may cause dwarfs with lower SFRs
$<0.1$ \Msun yr$^{-1}$ more likely to be missed in narrowband surveys.  On the other hand, the existence of
galaxies with high ${\mathcal M}$ implies that a high EW cannot unambiguously be interpreted as a defining characteristic of primordial population III galaxies \citep{2003A&A...397..527S,2009MNRAS.399...37J,2011ApJ...731...54P}. Finally, we use the framework presented in Dijkstra \& Wyithe 2012 to quantify the impact of stochasticity on the predicted luminosity function and EW-PDF of a sample of Ly$\alpha$ emitters (LAEs, i.e. narrowband selected galaxies) at a fixed redshift. We find that while stochasticity does not appreciably affect the LAE abundance modeling, it may help to explain the large EW (EW$\gsim 100-200$ \AA) observed for a fraction of LAEs.

Finally, we have also shown that stochasticity can affect inferred constraints on the escape fraction of ionizing radiation, most strongly in galaxies that form stars at rates SFR$\lsim 1 M_{\odot}$ yr$^{-1}$. Given that such galaxies are thought to dominate the reionization process at $z>6$, it is important to consider the effects of stochasticity when using `local' observations to put constraints on their ionizing luminosities and escape fractions. In particular, the observed anomalies in the ratio of LyC to non-ionizing UV radiation in so-called `Lyman-Bump' galaxies at $z=3.1$ reported by \cite{2009ApJ...692.1287I} can be explained naturally in the context of the stochastic sampling of the IMF, without resorting to unusual stellar populations. Moreover, the absence of such objects in a sample of $15$ LBGs at $z\sim 1.3$ \citep{2010ApJ...723..241S} is also expected as these galaxies are forming stars at rates at which the stochastic sampling of the IMF becomes negligible.

\section*{Acknowledgments} 
JEF-R acknowledges support from the Peter and Patricia Gruber Foundation through their Fellowship, administered by the International Astronomical Union. We thank Diederik Kruijssen for useful discussions on stellar clusters and IMF sampling.

\bibliographystyle{mn2e}

\end{document}